\documentclass[twocolumn,a4paper,10pt,twoside]{article}

\usepackage{amsmath}
\usepackage{amssymb}
\usepackage{fancyheadings}
\usepackage[dvips]{color}
\usepackage[dvips]{graphicx}

\newcommand{\pacz}{Paczy\'nski }

\newcommand{\maketitlesven}{
\begin{minipage}[t]{\textwidth}
\begin{center}
{\LARGE {\bf AGAPE : Gravitational microlensing effect to look}}\\
\vspace{0.2cm}
{\LARGE {\bf for MACHOs towards M31}}\\
\vspace{0.2cm}
{\bf St{\'e}phane Paulin-Henriksson}\footnote{PCC, Coll{\`e}ge de France, see below}\\
{\bf on behalf of POINT-AGAPE collaboration}\footnote{POINT-AGAPE is a collaboration between :\\
$\bullet$ Astronomy Unit, School of Mathematical Sciences, Queen Mary \& Westfield College, Mile End Road, London \mbox{E1 4NS}, UK\\
$\bullet$ Institute of Astronomy, Madingley Road, Cambridge \mbox{CB3 0HA}, UK.\\
$\bullet$ CERN, 1211 Gen{\`e}ve, Switzerland\\
$\bullet$ Laboratoire d'Astrophys. UMR CNRS 5572, Observatoire Midi-Pyr{\'e}n{\'e}es, 14 Avenue Edouard Belin, F-31400 Toulouse, France\\
$\bullet$ Laboratoire de Physique Corpusculaire et Cosmologie, Coll{\`e}ge de France, 11 Place Marcelin Berthelot, F-75231 Paris, France\\
$\bullet$ Universit{\'e} Bretagne-Sud, campus de Tohannic, BP 573, F-56017 Vannes Cedex, France\\
$\bullet$ Theoretical Physics, Oxford University, 1 Keble Road, Oxford \mbox{OX1 3NP}, UK}
\end{center}
\end{minipage}\hfill
}

\begin{document}

\thispagestyle{fancy}
\cfoot[\fancyplain{}{\bfseries\thepage}]
{\begin{minipage}{0.9\textwidth}
\begin{center}
\begin{tabular}{p{10cm}}
\\
\hline
\end{tabular}
\end{center}
{\bf  To appear in the proceedings of the \emph{Dark Univers} conference,}\\
{\bf  2 - 5 april 2001, Baltimore (\textsf{http://ntweb.stsci.edu/SD/DarkUniverse}). ed. Mario Livio.}
\end{minipage}}

\twocolumn[ \maketitlesven
\section*{Abstract}
We report the discovery of a short-duration microlensing candidate in the Northern field of the POINT-AGAPE pixel lensing survey towards M31 (\mbox{$\alpha = 00^{\textrm{h}}42^{\textrm{m}}51.42^{\textrm{s}}$} and \mbox{$\delta = +41^o23'53.7''$}). The Full-Width at Half-Maximum (FWHM) timescale is very short (\mbox{$t_{\textrm{{\footnotesize fwhm}}} \simeq 1.8$ days}) and the source star has almost certainly been identified on HST archival images, allowing to infer an Einstein crossing \mbox{time : $t_E = 10.4$ days} and a magnification at \mbox{maximum : $A_{\textrm{{\footnotesize max}}} \sim 18$}. 
\vspace*{0.6cm}
]

\section{Introduction}
\begin{figure}[h]
\begin{center}
\includegraphics[scale=0.4]{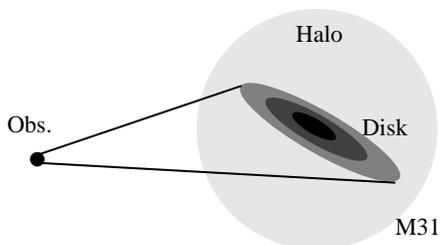}
\caption{The high inclinaison of M31's disk causes an asymmetry in the observed rate of microlensing events by lenses in a spheroidal halo \cite{crotts1}.}
\label{fig:inclinaison}
\end{center}
\end{figure}

In 1986, \mbox{B. \pacz \cite{paczynski}} suggested to search for \mbox{MACHOs} using gravitational microlensing on background stars. The possibility of detecting such events in M31 was independently suggested by \mbox{A. Crotts \cite{crotts1}} and \mbox{P. Baillon} et \mbox{al. \cite{baillon}}. The advantages of targeting such a large external galaxy are :\\
$\bullet$ the number of stars that can act as possible sources is enormous.\\
$\bullet$ this gives access to a different direction than \mbox{Magellanic} clouds to study the halo of the Milky Way.\\
$\bullet$ the high inclinaison of M31's disk causes an asymmetry in the observed rate of microlensing by lenses in a spheroidal halo, as shown in figure \ref{fig:inclinaison}. This gives an unambigous signature of the halo \mbox{lenses \cite{crotts1}}.

The main difficulty of such observations is that the sources are not resolved (nevertheless they might be resolved when they are lensed). The AGAPE \mbox{collaboration \cite{ansari1} \cite{agapez1}} has developped an analysis method called ``Pixel method'' (see \mbox{section \ref{sec:pixelmethod}}) and identified several candidate microlensing events. In particular two short-duration candidates was \mbox{reported : } \mbox{AGAPE-Z1 \cite{agapez1}} and \mbox{PA-N1 \cite{agapen1}}. The Columbia-VATT \mbox{group \cite{crotts2}} developped independently a method based on differential image analysis.
As in most cases we do not have acces to the magnitude of the sources, there is a parameter degeneracy in the theoretical ``\pacz fit'' and the way to compute the optical depth is more complicated than for resolved stars (see section \ref{sec:opticaldepth}).

\section{Pixel lensing}
\label{sec:pixellensing}

\subsection{Pixel method}
\label{sec:pixelmethod}

\begin{figure}[h]
\begin{center}
\includegraphics[scale=0.6]{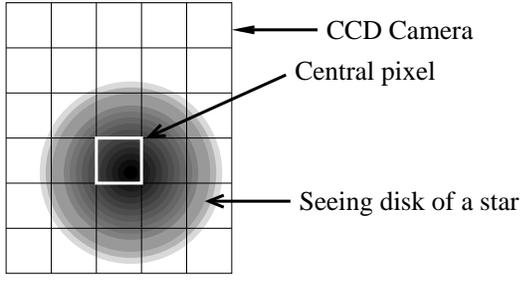}
\caption{The pixel method consists in the survey of the flux in each pixel. An event is detected is the increase of the flux in the central pixel is typically bigger than the noise : $\Delta \Phi_{^{\textrm{{\footnotesize central}}}_{\textrm{{\footnotesize pixel}}}} \gtrsim \sigma$.}
\label{fig:pixelmethod}
\end{center}
\end{figure}

The Pixel method was suggested by \mbox{P. Baillon} et \mbox{al. \cite{baillon}} and \mbox{A. Gould \cite{gould}}. The light-curve of every pixel is studied to detect significant variations. Due to the seeing, the flux of a star is spread over several pixels, as shown on \mbox{figure \ref{fig:pixelmethod}}. 
Therefore several neighbouring pixels are simultaneously affected by a single variation. To get the best signal to noise ratio it is necessary to find the central pixel of an event. The event is detected if the increase of the flux in the central pixel is typically bigger than the noise :
\begin{displaymath}
\Delta \Phi_{^{\textrm{{\footnotesize central}}}_{\textrm{{\footnotesize pixel}}}} = \left[ A-1\right]\times f \times \Phi_* \quad \gtrsim \quad \sigma
\end{displaymath}
where $A$ is the magnification, $f$ the seeing fraction and $\Phi_*$ the flux of the source. As only large amplifications are detected, the detection efficiency is weak but this is compensated by the huge number of monitoring sources.

\subsection{Optical depth in pixel lensing}
\label{sec:opticaldepth}
\begin{figure}[h]
\begin{center}
\includegraphics[scale=0.4]{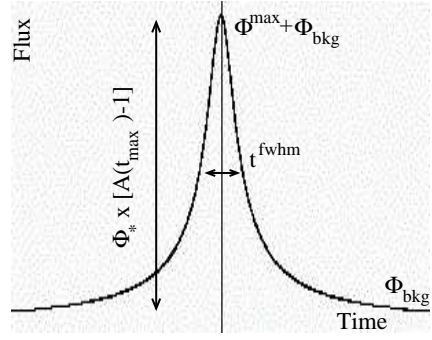}
\caption{In pixel lensing one can usually determine $\Phi_{\textrm{{\footnotesize bkg}}}$ and $\Phi_{\textrm{{\footnotesize max}}} = \Phi_* \times \left[ A(t_{\textrm{{\footnotesize max}}})-1\right]-\Phi_{\textrm{{\footnotesize bkg}}}$ but not $\Phi_*$ and so not $t_E$.}
\label{fig:degeneracy}
\end{center}
\end{figure}

The flux increase in a microlensing event is given by (for pointlike lens and pointlike source) \cite{einstein} :
\begin{eqnarray}
\Delta\Phi(t) & = & \Phi_* \times f(u^2(t))\nonumber\\
f(x) & = & \frac{2+x}{\sqrt{x(4+x)}}-1\nonumber
\end{eqnarray}
$\Phi_*$ is the source flux in absence of lensing, \mbox{$u(t)=\theta (t)/\theta_E$}, $\theta(t)$ is the angular separation of lens and source, and $\theta_E$ is the angular Einstein radius :
\begin{displaymath}
\theta_E=\sqrt{\frac{4GM}{c^2}\frac{s-l}{sl}}
\end{displaymath}
where $M$ is the lens mass and $s$ and $l$ the distances of the source and the lens from the observer. The relative motion of the lens and the source is usually well approximated by a uniform motion :
\begin{displaymath}
u^2(t) = \beta^2+\left( \frac{t-t_{\textrm{{\footnotesize max}}}}{\beta \quad t_E} \right)^2
\end{displaymath}
$t_{\textrm{{\footnotesize max}}}$ is the time of the maximum magnification, $\beta$ the impact parameter in units of the Einstein radius and $t_E$ is the Einstein time. In the limit of large magnifications at maximum, the microlensing curve assumes the degenerated form :
\begin{displaymath}
\Delta\Phi(t)\simeq \frac{\Phi_*}{u(t)} = \frac{\Phi_*}{\beta}\left[1+\left(\frac{t-t_{\textrm{max}}}{\beta t_E}\right)^2\right]^{-1/2}
\end{displaymath}
The three parameters $\Phi_*$, $\beta$ and $t_E$ have reduced to two: ``$\Phi_*/\beta$'' and ``$\beta t_E$''. A fit to the data can determine neither the star flux $\Phi_*$ nor the Einstein time $t_E$ separately from the unknown impact parameter $\beta$. This degeneracy is illustrated in figure \ref{fig:degeneracy}.

Usually, the estimated optical depth is defined as :
\begin{displaymath}
\overline{\tau} = \frac{\pi}{2N_sT}\sum_{^{\textrm{{\footnotesize detected}}}_{\textrm{{\footnotesize events}}}} \frac{t_{E_i}}{\epsilon_i}
\end{displaymath}
where $\overline{\tau}$ is the estimated optical depth, $N_s$ the number of observed sources, $T$ the duration of observations, $t_{E_i}$ the Einstein time of event $i$ and $\epsilon_i$ the detection efficiency for event $i$. As it is shown above, it is usually not possible to determine $t_E$ in pixel lensing. Remarkably, however,  it is it still possible to evaluate the optical depth. P.Gondolo \cite{opticaldepth} shown that the estimated optical depth can be written :
\begin{displaymath}
\overline{\tau} = \frac{0.72}{TF_{\textrm{{\footnotesize eff}}}(\Phi^{\textrm{{\footnotesize max}}}_{\textrm{\footnotesize lim}})}\sum_{\Phi^{\textrm{{\footnotesize max}}}_i > \Phi^{\textrm{{\footnotesize max}}}_{\textrm{\footnotesize lim}}} t^{\textrm{{\footnotesize fwhm}}}_i \Phi^{\textrm{{\footnotesize max}}}_i
\end{displaymath}
where $\Phi^{\textrm{{\footnotesize max}}}_{\textrm{\footnotesize lim}}$ is the detection threshold and $F_{\textrm{{\footnotesize eff}}}$ a function of $\Phi^{\textrm{{\footnotesize max}}}_{\textrm{\footnotesize lim}}$ which depends slightly on the luminosity function of sources.

\section{POINT-AGAPE collaboration}
The POINT-AGAPE collaboration uses the Wide Field Camera (WFC) on the 2.5 m Isaac Newton Telescope \mbox{(INT) \cite{int}} at La Palma, Canarie Islands . It carries out a pixel-lensing survey, monitoring two fields of \mbox{0.3 deg$^2$} each, located North and South of the M31 center. The pixel size is 0.33$''$. The survey has the potential to map the global distribution of the microlensing events in M31 and to determine any large scale gradient. Three years of data are planned (from August 1999 to January 2002).

\subsection{Data analysis}
\begin{figure}[h]
\begin{center}
\includegraphics[scale=.4]{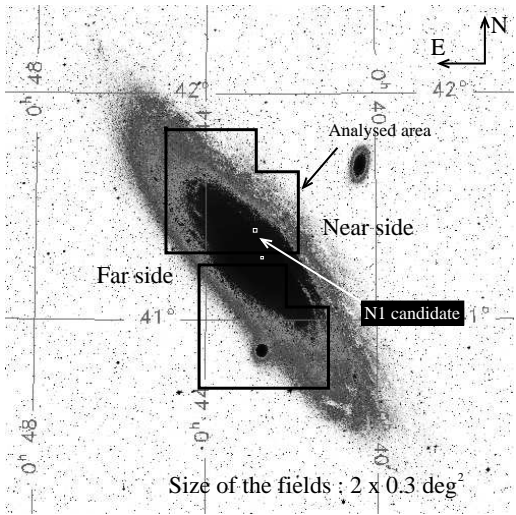}
\caption{The two fields observed by POINT-AGAPE collaboration}
\label{fig:fields}
\end{center}
\end{figure}

For the moment we restrict the analysis to the Northern field in 1999, as shown figure \ref{fig:fields}. The observations are spread over 36 epochs between August and December. The exposures are in two bands : 36 nights in \mbox{Sloan r$'$} and 26 in \mbox{Sloan g$'$}. The exposure time is typically between 5 and 10 minutes per night and per band.

After bias subtraction and flat-fielding, each image is geometrically and photometrically aligned relative to a reference image (14 August 1999) which was chosen because it has a long exposure time, typical \mbox{seeing (1.6$''$)} and little contamination from the Moon. The light curves are made for ($2.3''\times 2.3''$)-superpixels to cope with seeing variations\footnote{This size is set by the worst seeing ($\sim$2.1$''$).}.

We use a simple set of selection criteria designed to isolate high signal-to-noise events. Detection of events is made in the Sloan r$'$ band which has better sampling and lower sky background variability. The Sloan g$'$ band is used to check achromaticity.

\subsection{Preliminary results}
In 1999 data and in Northern field, we have already found about 300 microlensing candidates (compatible with a \pacz{} fit) which will have to be validated using 2000 and 2001 data. Magnitudes at maximum of amplification are in the \mbox{range :} \mbox{$21 \lesssim \textrm{R}_{\textrm{max}} \lesssim 24.5$}, where R is the standard Johnson-Cousins band.

One of our candidates, called PA-N1, is particularly interesting with \mbox{$t_{\textrm{fwhm}} < 2$ days} and \mbox{$\Delta\Phi_{\textrm{max}} \sim 18$ ADU.s$^{-1}$}, corresponding to $R\sim 20.8$. This candidate, presented \mbox{in \cite{agapen1}}, is the subject of the next section.

\section{PA-N1 microlensing candidate}
\label{sec:n1}
\begin{figure*}[ht!]
\centering
\includegraphics[scale=0.65]{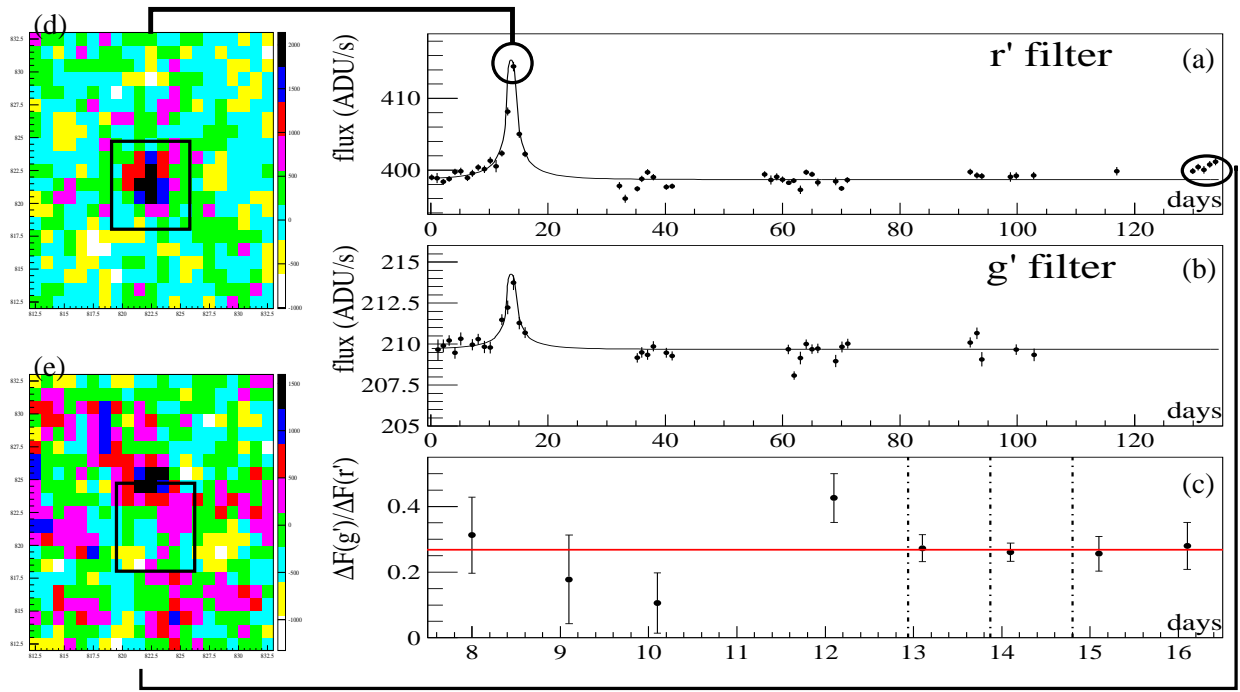}
\caption{\textbf{Panels a and b :} flux in r$'$ and g$'$ against time in days; \mbox{days = J-2451392.5} where J is the Julian date, i.e: data are spread over August and December 1999. \mbox{\textbf{Panel c :}} zoom centered on the event showing the variation of the ratio of the flux change in the two passband with time. The vertical lines are centred on $t_{\textrm{max}}$ and are separated by 0.9 days, i.e. 1/2 FWHM. \mbox{\textbf{Panels d and e :}} positions (using image differencing) of PA-N1 and the December bump. One can see that the two sources are clearly separated \mbox{by $\gtrsim 3$ pixels}, \mbox{i.e $\sim 1.1''$} (pixel size is $0.33''$). However the December source is close enough to PA-N1 to be visible on the superpixel lightcurve.
}
\label{fig:n1}
\end{figure*}

Figure \ref{fig:n1} (panels a and b) shows the lightcurves in \mbox{Sloan r$'$} and \mbox{Sloan $g'$} of this candidate together with the theoretical fit derived below. 
\begin{table*}[ht!]
\begin{center}
\begin{tabular}{|c|c|c|c|c|c|c|}
\hline
$t_{\textrm{max}}$ & $\beta$ & $t_E$ & $\Phi_{*,r'}$ & $\Phi_{*,g'}$ & $\Phi_{\textrm{bkg},r'}$ & $\Phi_{\textrm{bkg},g'}$ \\
(days) & & (days) & (ADU/s) & (ADU/s) & (ADU/s) & (ADU/s)\\
\hline
13.87 & 0.056 & 10.4 & 1.02 & 0.28 & 397.65 & 209.4\\
\hline
\end{tabular}
\begin{displaymath}
\Rightarrow A_{\textrm{max}} \simeq 18
\end{displaymath}
\caption{Fit parameters assuming the amplified star is the red giant visible on HST images.}
\label{tab:paramn1}
\end{center}
\end{table*}
Using the Aladin Sky \mbox{Atlas \cite{aladin}}, we found that PA-N1 has J2000 \mbox{position :} \mbox{$\alpha = 00^{\textrm{h}}42^{\textrm{m}}51.42^{\textrm{s}}$} and \mbox{$\delta = +41^o23'53.7''$}. That is, it lies at $7'52''$ from the center of M31. On images close to the maximum amplification, PA-N1 is bright enough to perform its photometry. We \mbox{found :} 
\mbox{$V=22.00\pm 0.17$} and \mbox{$R=20.87\pm 0.13$} \mbox{$\Rightarrow$ $V-R=1.2\pm 0.22$} which is very red.
On figure (panel a) one can see that, after \mbox{$t \sim 120$ days} (\mbox{i.e in December} 1999), a second bump appears on the superpixel light curve. Using image subtraction, one can show (panels d and e) that this second bump is not due to the same source as the first one. Distance between the two sources is about $1.1''$. December points were excluded from the fit.

\begin{figure}[h]
\begin{center}
\includegraphics[scale=0.4,clip=]{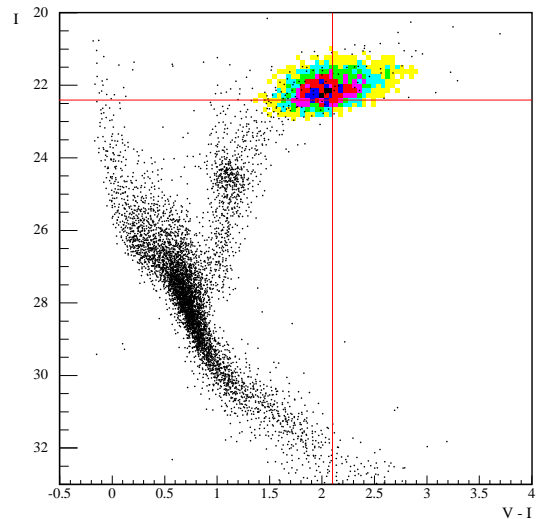}
\caption{Superposition of the Hipparcos (points) and resolved stars in HST images (shaded area) color-magnitude diagrams. The Hipparcos stars was moved on the position on M31. The cross hairs show the position of the HST source (\mbox{$V=24.51\pm 0.12$} and \mbox{$V-I=2.10\pm 0.16$}).}
\label{fig:diaghr}
\end{center}
\end{figure}

The PA-N1 position lies within a serie of five \emph{Hubble Space Telescope} (HST) WFPC2 archival images taken in July 1996, three with F814W filter \mbox{($\sim$ I)} and two with F606W filter \mbox{($\sim$ V)}. On these images we found a resolved source close to the \mbox{$1 \sigma$ error} box and no other within the \mbox{$3\sigma$ error} box. The prior probability to find such a star so close to the predicted position is only 3\%. The color of this resolved star (\mbox{$V=24.51\pm 0.12$} and \mbox{$V-I=2.10\pm 0.16$}) is perfectly compatible with the color of PA-N1. Moreover \mbox{figure \ref{fig:diaghr}} 
shows that the source star is either the star visible on HST images or must be on the main sequence and so magnified by $A_{\textrm{max}} \gtrsim 10^4$. The latter possibility is extremely unlikely. Hence a firm prediction of the microlensing interpretation is that the source is almost certainly identified to be the red giant visible on HST images. Using this information, we can eliminate the degeneracy between fit parameters. The parameters are shown in table \ref{tab:paramn1}.

\pagebreak

\end{document}